\documentclass[conference]{IEEEtran}
\usepackage{cite}
\usepackage{amsmath,amssymb,amsfonts}
\usepackage{algorithmic}
\usepackage{graphicx}
\usepackage{textcomp}
\usepackage{xcolor}
\def\BibTeX{{\rm B\kern-.05em{\sc i\kern-.025em b}\kern-.08em
    T\kern-.1667em\lower.7ex\hbox{E}\kern-.125emX}}

\usepackage{multirow}
\usepackage{comment}
\usepackage{subfigure}

\usepackage{balance}

\graphicspath{{pics/}}
    
\begin{document}
\bstctlcite{BSTcontrol}

\title{Intention Detection of Gait Adaptation in Natural Settings\footnotemark{\textsuperscript{1}\\
} 

}


\author{
\IEEEauthorblockN{Ines Domingos}
\IEEEauthorblockA{\textit{Hamlyn Centre} \\
\textit{Imperial College London}\\
London, United Kingdom \\
ines.domingos16@imperial.ac.uk}
\and
\IEEEauthorblockN{Guang-Zhong Yang}
\IEEEauthorblockA{\textit{Medical Robotics} \\
\textit{Shanghai Jiao Tong University}\\
Shanghai, China \\
gzyang@sjtu.edu.cn}
\and
\IEEEauthorblockN{Fani Deligianni}
\IEEEauthorblockA{\textit{School of Computing Science} \\
\textit{University of Glasgow}\\
Glasgow, United Kingdom \\
fani.deligianni@glasgow.ac.uk}

}

\maketitle
\addtocounter{footnote}{1}
\footnotetext
{Authors acknowledge funding from EPSRC EP/R026092/1 and EP/R045178/1 \\
Ethical approval has been obtained from ICREC 18IC4816}

\begin{abstract}
Gait adaptation is an important part of gait analysis and its neuronal origin and dynamics has been studied extensively. In neurorehabilitation, it is important because it enables neuroplasticity mechanisms and facilitates the restoration of motor function. For this reason, brain–computer interfaces (BCI) have been build to facilitate neurorehabilitation. 
This paper presents a gait adaptation scheme in natural settings. It allows monitoring of subjects in more realistic environment without the requirement of specialized equipment such as treadmill and foot pressure sensors. We extract gait characteristics based on a single RGB camera whereas wireless EEG signals are monitored simultaneously. Based on Regularised Common Spatial Patterns (RCSP) that take into consideration both amplitude and frequency EEG features, we demonstrate that the method can not only successfully detect adaptation steps but it also detect efficiently whether the subject adjust their pace to higher or lower speed. 
\end{abstract}


\section{Introduction}

Gait adaptation involves the ability to change walking direction and/or speed to avoid obstacles and keep balance. Deficiency in walking adaptation indicates a risk factor of falling in the elderly population or patients with Parkinson or stroke \cite{Wagner,Gwin,Martelli}. 
Gait adaptation plays an important role in neurorehabilitation since it perturbs neuronal dynamics and allows patients to restore motor function. Predictive frameworks of gait adaptation may reveal suitable interventions for an effective gait rehabilitation. Additionally, intention detection of movement and gait adaptation is a successful way to integrate a lower limb robotic system in patient’s rehabilitation.

Following a neurological injury, such as stroke or spinal cord injury, the key to gait recovery, is neuroplasticity, which is an activity-dependent change in brain structure and function. For example, repetitive motion patterns enhance neuronal connections involved in the underlying motor task but they could also trigger suboptimal compensation mechanism \cite{Jones,Edgerton}. Therefore, the timing of initiating therapeutic exercises and movements as well as the content of the exercises are of paramount importance. 

Numerous researches have shown that auditory rhythm has a deep effect on the motor system. These studies show that there is a strong connectivity across cortical, subcortical, and spinal levels between the auditory and motor systems \cite{Thaut}. Based on these evidences, some gait adaptation studies focus on auditory rhythms, where patients try to couple heel strikes and pacing tones, improving the gait coordination. Consequently, gait adaptation based on split-zone treadmill exercises and auditory rhythm has shown to improve gait symmetry in patients with stroke, cerebral palsy and Parkinson disease \cite{Thaut,Roerdink1,Howe} and is an effective way to adapt stride frequency and improve gait coordination in people after stroke \cite{Roerdink1}.

Here we study gait adaptation based on a rhythmic tone that alternates between three modes of slow, normal and fast pace. The subjects follow the tone as they walk inside a room without any further restriction. The EEG signal is simultaneously recorded via wireless devices. Contrary to previous studies we do not use a treadmill or specialized equipment, which allows the investigation of gait adaptation in more natural settings. We capture gait characteristics such as heel strikes based on a single RGB camera. Subsequently, we use this information and behavioral analysis of the reaction time to extract gait adaptation steps versus non-gait adaptation steps.

We preprocess the EEG signal based on bandpass filtering and independent component analysis (ICA) to remove motion related artefacts and subsequently the signal is epoched based on right/left heel strikes. Finally, EEG gait adaptation characteristics are investigated based on three classification problems: i) right versus left gait cycle classification (two classes); ii) adaptation versus non adaptation steps (two classes) and iii) adaptation to higher pace versus adaptation steps towards lower pace versus non adaptation steps (three classes). To this end, we extract features based on regularized common spatial patterns (RCSP) and instantaneous frequency estimation based on the Hilbert transform. Our results show that we can successfully discriminate adaptation versus non-adaptation with more than 90\% testing accuracy, which corresponds to 0.06 10-fold cross-validated generalization loss. Furthermore, we show that combining amplitude and frequency characteristics of the EEG signal outperforms each of them alone.

\section{Related Work}

\subsection{BCI in gait rehabilitation}
A growing number of studies investigates brain activity during human locomotion with EEG data. Preceding studies found that cerebral activity increases during walking or preparation for walking and there is a significant activation of the sensorimotor area, during isolated leg or foot movements as well as during gait. 
 It is also believed that neuronal activity has different functional roles according to the frequency ranges, which provide finer details on which brain network features are important in gait control \cite{Seeber}. 
 
Recently, brain–computer interfaces have been used as a rehabilitation therapy to restore the motor functions in people with gait impairments. Specifically, there has been a huge interest in the use of BCIs in post-stroke gait therapy \cite{Ramos}.
This technology can be used in two different approaches. It can be used to control directly the rehabilitation devices or to provide feedback to the user based on brain activity. The feedback is provided by output of rehabilitation devices, for example, the movement of a prosthetic limb, activated with brain activity. Later, when brain activation associated with motor intention is measured the information is extracted and used as a signal to control external devices. For the purpose of BCI, the better neural control signal is found in the range of 8-13 Hz ($\mu$-rhythm), which is found in the central sensory-motor areas. 

\subsection{Regularised Common Spatial Patterns}
CSP has been used successfully before for feature extraction in gait experiments \cite{Zhang3}. CSP is known to be very popular and effective but it is also affected by noise and may overfit, especially with small datasets \cite{Lotte}. To overcome these disadvantages of the CSP method, there has been a vast interest in adding prior information to the CSP learning process, using regularization terms \cite{Lu,Kang,Ledoit}. The process of adding prior information into the CSP method can be achieved with two distinct manners. It can be done either at the covariance matrix estimation or at the level of the objective function, which imposes prior information on the spatial filters \cite{Lotte}. 

So far, several RCSP algorithms were developed; among the most notable, the composite CSP (CCSP), the regularized CSP with generic learning approach and the regularized CSP with diagonal loading. The goal of the composite CCSP, proposed by Kang et al \cite{Kang}, is to perform subject-to-subject transfer, which regularizes the covariance matrix using other subjects’ data. The Regularized CSP with generic learning approach was proposed by Lu et al \cite{Lu} and aims to regularize the covariance matrices using data from other subjects. The Regularized CSP with diagonal loading approach uses the Ledoit and Wolf’s method to  decrease the covariance matrix towards the identity matrix \cite{Ledoit}. 
Usually, these approaches are based on the amplitude of the EEG signal and ignore its frequency characteristics.

\section{Methods}

\begin{figure*}[t]
    \centering

        \includegraphics[width=.99\linewidth]{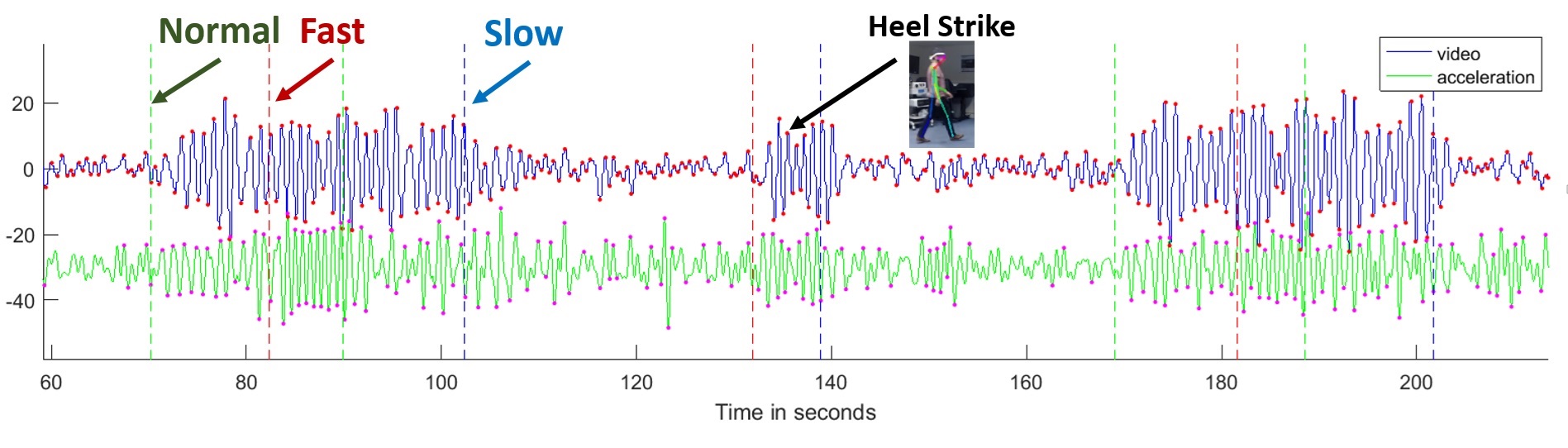}

    \caption{Gait features extraction: The video-based gait signal and the acceleration-based gait signal are plotted in blue and green lines, respectively after SSA processing. The extrema points represent left and right heal strikes, respectively. Adaptation events are represented as vertical lines in green, red and blue colors that represent, normal, fast and slow speed, respectively.}
     \label{fig:overview}
\end{figure*}

\subsection{Experimental Setup and Procedures}
EEG data was recorded from six healthy participants ($27.5 \pm 7.58$ years). A 32-channels, g.tec Nautilus, EEG wireless acquisition system with active-electrodes (Ag/AgCl) was used. The system records EEG data at 250Hz along with acceleration data in three axes. The EEG cap was placed accordingly to the 10-20 system. Impedance was measured to ensure a value of less than $30\Omega$ for all participants. Data were recorded with openvibe version 1.3 \cite{Lotte}. 

Participants were asked to walk according to a musical tone that it was programmed to switch between three modes, slow walking, normal walking and fast walking. The frequency at normal walking was 1.75Hz, whereas at slow walking speed was halved and in fast walking, speed was 1.5 times the normal. The duration of each mode was estimated to be around $24\pm 6$ right/left steps, which expresses a random variation of up to six steps. Each mode consisted of 20 trials, which resulted in a total of 60 adaptations randomly permuted. The overall experiment lasts about 16 minutes. The stimulus was programmed and displayed with Psychtoolbox-3 \cite{Brainard,Pelli,Kleiner}. Adaptation events were send to the EEG acquisition server via TCP/IP communication.

A Logitech camera has been also used to record participants at 60 frames per second while they were walking. In order to synchronize the camera recording with the EEG acquisition, each captured frame raised an event that was send to the EEG acquisition server via TCP/IP communication. Video capturing and events’ transmission was also implemented with Psychtoolbox-3.

\subsection{Gait Features Extraction}
We are interested in epoching the EEG signal into segments according to left/right heel strikes. Towards this aim, we obtain gait information based on the camera recordings and the acceleration data of the EEG system. To detect and track 2D coordinates of human joints based on a single RGB camera, we used OpenPose \cite{Cao,Wei,Simon}. This is a state-of-the-art, real-time approach that uses deep neural networks to track the joints of multiple-persons stably. 

Gait analysis based on a single RGB camera is challenging due to the perspective projection and limited 2D information \cite{gu_markerless_2018,deligianni_emotions_2019,guo_3-d_2019,Gu2021}. To extract gait events of right and left heel strikes, we estimate the Euclidean distance between the left and right ankle coordinates in Y-camera axis, assuming that the camera is in a vertical position. Singular spectrum analysis (SSA) has been applied to denoise the signal and improve the detection of peaks that reflect foot contacts. SSA has been also used, successfully, to detect heel strikes based on acceleration data \cite{Jarchi,Deligianni1}.
SSA is based on time-series subsampling to construct a trajectory matrix, the so called Hankel matrix. If $w$ is a time-series, then the trajectory matrix takes the form:

\begin{equation}
\begin{split}
W=
\begin{pmatrix}
w_0 & \cdots & w_{p-1}\\
\vdots & \ddots & \vdots \\
w_{l-1} & \cdots & w_{m-1}
\end{pmatrix}
\end{split}
\end{equation}

Where $p=m-l+1$, $m$ is the length of $w$ and $l$ is the embedding dimension. The signal is reconstructed from averaging of a subset of the group elementary matrices of the decomposition of the covariance matrix: $C_W=WW^T$.

Here we process acceleration data that comes with the gtec acquisition system to ensure that the EEG signal and the video timeline is fully synchronized. Acceleration data are processed with Principal Component Analysis (PCA) to derive the dominant signal variation, which is due to gait. Singular spectrum analysis (SSA) and peak detection was also utilized to detect heel strikes. An example of gait features' extraction from both the video and the accelerometer is shown in Figure \ref{fig:overview}.


\section{Movement Artefact Removal}
EEG-data acquisition is very sensitive to motion artefacts. To eliminate the influence of motion, we have filter the EEG signal based on a bandpass impulse response (FIR) filter of 3-45Hz. The filter is applied forward and then backward to ensure that phase delays are eliminated.  

Subsequently, we use independent component analysis (ICA) based on the infomax algorithm to remove the influence of motion components \cite{Makeig}. ICA is a common approach of removing gait-related movement artefacts \cite{Snyder,Gwin}. It involves the extraction of maximally independent components. Motion components are normally identified manually based on their frequency profile and their spatial distribution. Subsequently, they are removed and the EEG signal is reconstructed without their influence.

\subsection{Feature Extraction based on Common Spatial Patterns}

Here, we investigate gait adaptation by formulating a classification problem of whether a step is an adaptation step or not. We identify adaptation steps based on the reaction time (RT) between the change of the rhythmic tone and the step to match the average step of the session. Non-adaptation steps are drawn from the middle of the trial to match the number of the adaptation steps.  

To extract classification features from the EEG data, we use the Common Spatial Patterns (CSP) algorithm, which extracts spatial filters that maximize the discriminability between two classes \cite{Lotte}.  CSP uses spatial filters s that maximize the following equations:

\begin{equation}
J(s)=\frac{s^{T}X_1^{T}X_1s}{s^{T}X_2^{T}X_2s}=\frac{s^{T}C_{1}s}{s^{T}C_{2}s}
\end{equation}

$X_i$ denotes the matrix $k\times n$ for class $i$, where $k$ is the number of samples and $n$ is the number of channels. $C_i$ is the covariance matrix of the EEG signal from class $i$, assuming a zero mean.

This problem is transformed to a standard eigenvalue problem by noting that it is equivalent to maximizing the following function derived based on the Lagrange method \cite{Lotte}:

\begin{equation}
L(\lambda,s)=s^{T}C_{1}s-\lambda (s^{T}C_{2}s-1)
\end{equation}

Since we are looking for the extreme points of the function the derivate of $L$ with respect to  $s$ is zero and therefore:

\begin{equation}
C_2^{T}C_{1}s=\lambda s
\end{equation}

\subsubsection{Regularised Common Spatial Patterns (RCSP)}
Although, the CSP filters are an efficient way of extracting spatial filters that discriminate two classes, they are sensitive to noise and outliers. We have devised an automated way of extracting right/left heel strikes that occasionally suffer from erroneous peak detection. To minimize the influence of these outliers in extracting features based on the CSP algorithm we use regularization. 

We adopt the Ledoit and Wolf’s method, which regularizes the covariance matrix by shrinking it to identity \cite{Ledoit}. In other words, the goal is to find a linear combination of the identity matrix, $I$, and the covariance matrix, $C_i$, whose expected quadratic loss is minimum.

\begin{equation}
C_{i}^{*}=(1-\rho)C_i+\rho I
\label{eq:LW}
\end{equation}

This method has been shown to be particularly effective in problems where the number of samples is less than the number of dimensions \cite{Deligianni2,deligianni3}. One of its major advantage is that the shrinkage parameter is estimated automatically based on the intrinsic properties of the signal. 

\subsubsection{Common Spatial Patterns of the Hilbert Transform}
Along with the amplitude information of the EEG signal, we are interested in incorporating instantaneous frequency features extracted based on the Hilbert transform of the pre-processed EEG signal, which can be considered a real valued process $x(t)$ .  
The Hilbert transform is given by the equation:

\begin{equation}
x_h(t)=\frac{1}{\pi}PV\int_{-\infty}^{\infty} \frac{x(\tau)}{t-\tau}d\tau
\end{equation}
where $PV$ represents the Cauchy principal value of the integral. 
We estimate the phase of the signal as: 
\begin{equation}
\phi(t)=\arctan{\frac{\Im{x_h(t)}}{\Re{x_h(t)}}}
\end{equation}
where $\Im{x_h(t)}$ and $\Re{x_h(t)}$ is the imaginary and real part of the Hilbert transform of the EEG signal, respectively. The instantaneous frequency is estimated based on the derivative of the phase. 
Subsequently, we apply the common spatial patterns Eq. \ref{eq:LW} to the instantaneous frequency to extract frequency based characteristics.  

\subsubsection{From two-class to multi-class formulation}
We are interested in not only identifying the intention to adapt but also determining whether the adaptation is from a slower to faster pace or vice-versa. Therefore, we device a three-classes classification problem that includes adaptation to higher speed, adaptation to lower speed and non-adaptation. CSP and RCSP are intrinsically two-class methods. To overcome this problem, we construct three pairs of filters between each combination of the three classes.  

\section{Results}
In this study, the investigation of gait and gait adaptation is based on the following EEG classification experiments:
\begin{itemize}
\item Right vs left gait cycle classification (two classes). 
\item Adaptation vs non adaptation steps (two classes).
\item Adaptation to higher speed vs. adaptation steps towards lower speed vs. non adaptation steps (three classes).
\end{itemize}

\begin{table}[htbp]
\centering
\caption{Behavioral Analysis of Adaptation: Step Statistics}
\label{tab:steps}

\begin{tabular}{ |p{1.2cm}||p{1.8cm}|p{1.8cm}|p{1.8cm}|  }
 \hline

Adaptation  Type & Step duration (secs)  & Adaptation time (secs) & Adaptation Steps \\
 \hline
 Slow  & $0.57\pm 0.039$                & $1.12\pm 0.4$                 & $1.95\pm 0.51$  \\
 Normal & $0.56\pm 0.001$                & $1.91\pm 0.26$                 & $3.36\pm 0.4$ \\
 Fast & $0.46\pm 0.031$                & $1.66\pm 0.056$                 & $3.63\pm 1.16$ \\
 
 \hline
\end{tabular}

\end{table}

Table \ref{tab:steps} summarises the behavioral analysis that includes the estimation of step time for each adaptation type: slow, normal and fast. It also includes the time and number of steps required to adapt from one condition to another. This is also called reaction time (RT) and it is estimated as the time between the change of the rhythmic tone and the time when the step matches the average step of the session within the standard deviation limit. We report a conservative RT, since we do not include the sessions where the end of adaption is not detected.

\begin{figure}[t]
    \centering

        \includegraphics[width=.99\linewidth]{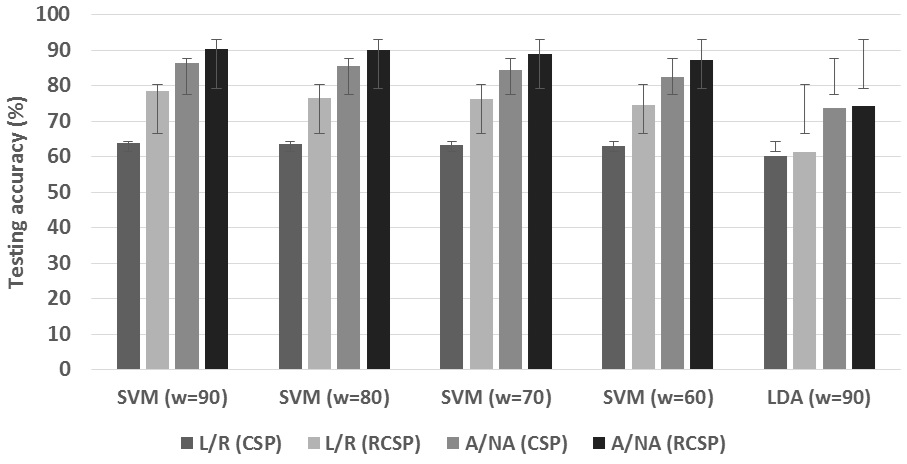}

    \caption{Testing accuracy of the classifications results of a) Left versus Right (L/R) steps and b) Adaptation versus non-adaptation (A/NA) steps. We compare results based on CSP and RCSP feature extraction. The results are shown across different sizes of sliding window (w).}
     \label{fig:accuracyVSmethod}
\end{figure}

\begin{figure}[t]
    \centering

        \includegraphics[width=.99\linewidth]{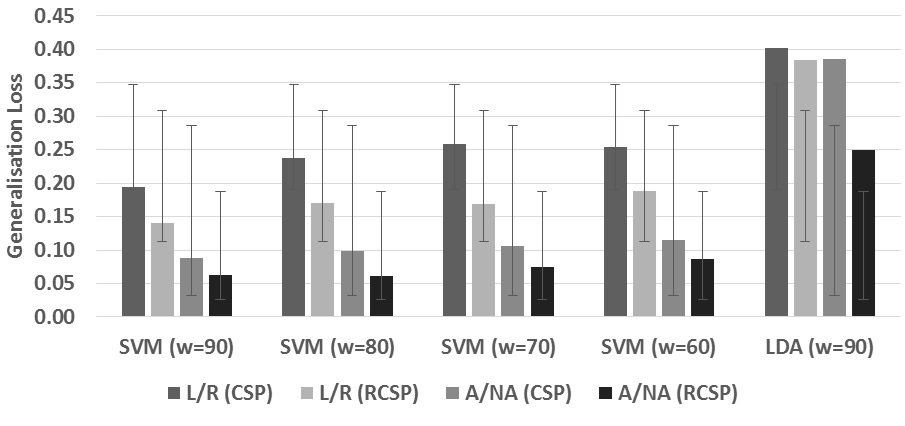}

    \caption{10-fold cross-validation generalization loss of the classifications results of a) Left versus Right (L/R) steps and b) Adaptation versus non-adaptation (A/NA) steps. We compare results based on CSP and RCSP feature extraction. The results are shown across different sizes of sliding window (w).}
     \label{fig:LossVSmethod}
\end{figure}

The classification results for the right versus left gait cycles and for the adaptation versus non-adaptation steps are summarized in Figure \ref{fig:accuracyVSmethod} and Figure \ref{fig:LossVSmethod}. Figure \ref{fig:accuracyVSmethod} demonstrates the testing accuracy of the classifications results of a) left versus right (L/R) steps and b) adaptation versus non-adaptation (A/NA) steps. We compare results based on CSP and RCSP feature extraction. The results are shown across different sizes of sliding window (w) with a range from 90 samples to 60 samples. Note that 90 samples correspond to 0.36 seconds, whereas the event duration is taken to be 0.4 seconds. In practice the step size may vary as it is shown in Table \ref{tab:steps}. For the adaptation versus non-adaptation training/testing set we assumed that the first three steps after each change of rhythmic tone are adaptation steps and subsequently we chose three steps at the middle of each adaptation trial as the non-adaptation steps. For classification we used support vector machines (SVM) based on radial basis function but we also show the results based on linear discriminant analysis (LDA).

Figure \ref{fig:LossVSmethod} demonstrates the 10-fold cross-validation generalization loss of the classifications results of a) left versus right (L/R) steps and b) adaptation versus non-adaptation (A/NA) steps. We compare results based on CSP and RCSP feature extraction. The results are shown across different sizes of sliding window (w), similarly to Figure \ref{fig:accuracyVSmethod}.

\begin{figure*}[t]
    \centering

        \includegraphics[width=.99\linewidth]{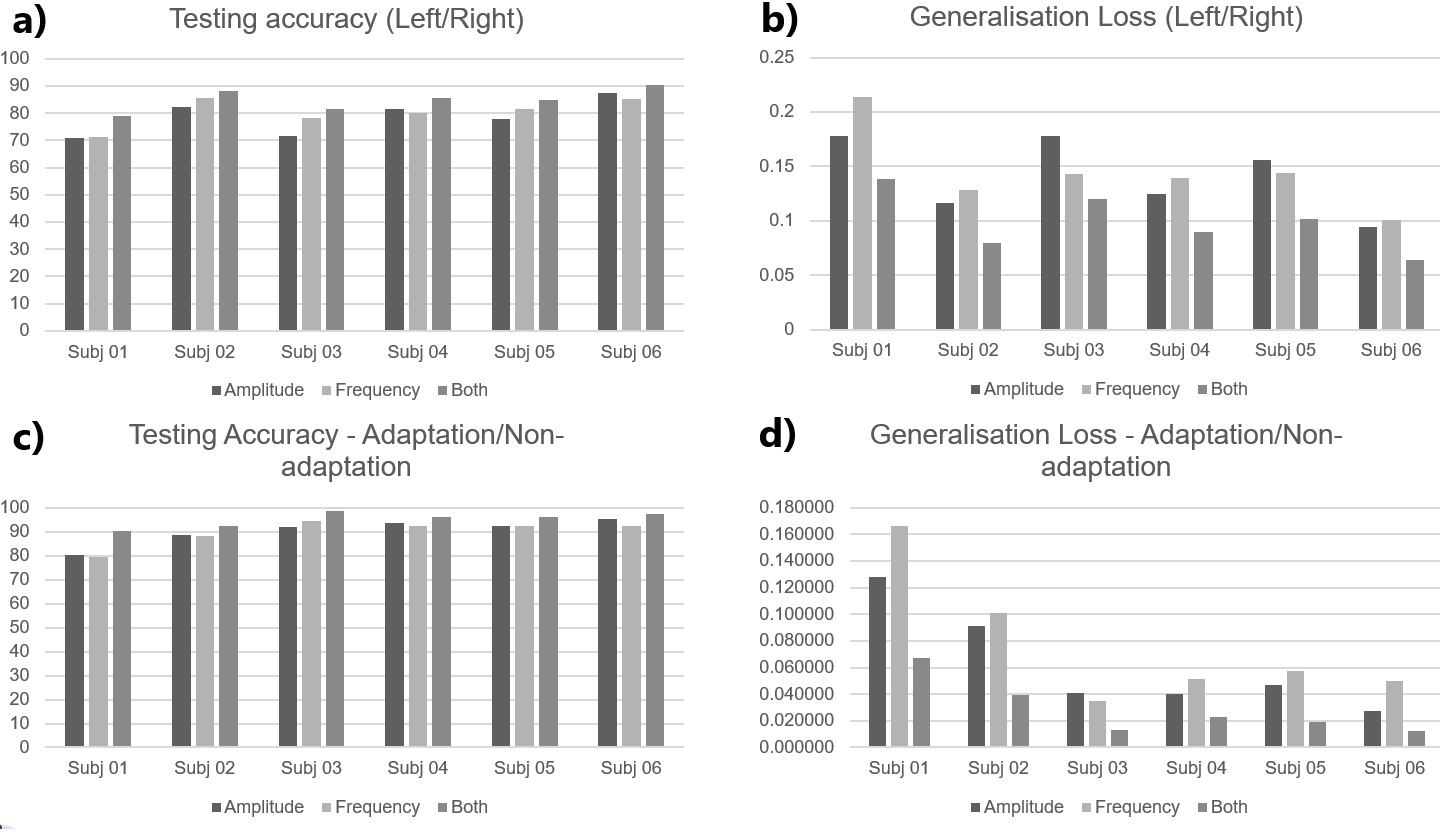}

    \caption{Application of common support patterns on both the amplitude and instantaneous frequency characteristics of the Hilbert transform of the signal: a) shows the testing accuracy for Left versus Right (L/R) steps, b) shows the generalisation loss for each subject for Left versus Right (L/R) steps, c) shows the the testing accuracy for each subject for adaptation versus non-adaptation (A/NA) steps, and d) shows the generalisation loss for each subject for adaptation versus non-adaptation (A/NA) steps.}
     \label{fig:frequency}
\end{figure*}

Figure \ref{fig:frequency} demonstrates the improvement of classification with the application of common support patterns on both the amplitude and instantaneous frequency characteristics of the Hilbert transform of the signal. In particular, Figure \ref{fig:frequency}a-b) shows the testing accuracy and generalisation loss for each subject, respectively, for Left versus Right (L/R) steps. 
Figure \ref{fig:frequency}c-d) shows the the testing accuracy and generalisation loss for each subject, respectively, for adaptation versus non-adaptation (A/NA) steps. We see a consistent improvement of the classification of performance across all subjects when both amplitude and frequency characteristics are incorporated. 

Table \ref{tab:confM} demonstrates the average confusion matrix (percentage) across subjects for the three-class classification problem. Feature extraction was based on three RCSP filters for each combination of classes. Classification was performed based on SVM with a linear kernel to avoid overfitting. Results are shown for a sliding window of 90 samples, which was shown to be most effective. The diagonal elements of the matrix represent the sensitivity results for each class. Note that the summation of the vertical columns results in 100\%.

\begin{table}[htbp]
\centering
\caption{Confusion Matrix (\%) For Three Class Classification}
\label{tab:confM}

  \begin{tabular}{|p{0.25cm} |p{1.75cm}|p{1.45cm}|p{1.45cm}|p{1.45cm}|}
 \hline
 & \multicolumn{4}{c|}{True Class} \\
\cline{2-5}

 \parbox[t]{2mm}{\multirow{4}{*}{\rotatebox[origin=c]{90}{predicted class}}} & Adaptation type & Slower to faster & Faster to slower & Non-Adaptation\\
 \cline{2-5}
 & Slower to faster & $84.79\pm 4.66$  & $5.08 \pm 3.32$ & $8.69 \pm 2.49$\\
 & Faster to slower & $6.34\pm 3.45$ & $86.73 \pm 3.36$ & $8.08 \pm 3.92$\\
 & Non-adaptation & $0.86 \pm 2.85$ & $8.17 \pm 2.32$ & $83.22 \pm 5.03$ \\
 \hline
 \end{tabular}
 \end{table}

\begin{figure}[t]
    \centering

        \includegraphics[width=.99\linewidth]{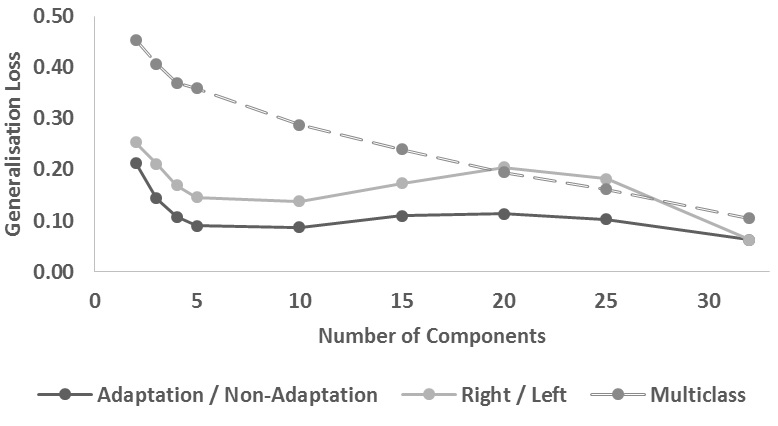}

    \caption{10-fold cross-validation generalisation loss across the most significant RCSP components for a sliding window of 90 samples.}
     \label{fig:LossVScomponents}
\end{figure}

\begin{figure*}[t]
    \centering

        \includegraphics[width=.99\linewidth]{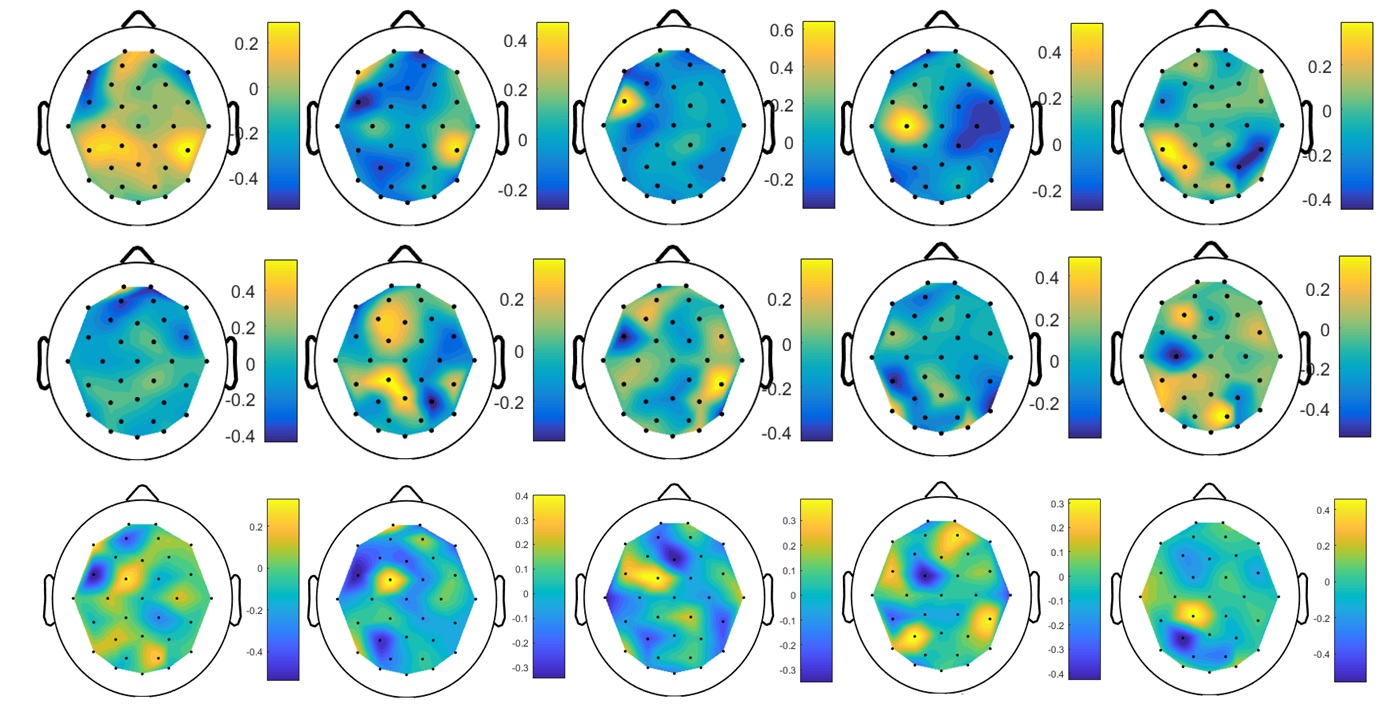}

    \caption{RCSP filters for the five most significant components of one of the subjects. a) Top row shows the RCSP filters associated with adaptation/non-adaptation classification, b) Middle row shows the RCSP filters associated with right/left classification, c) Bottom row shows the RCSP filters associated with the adaptation to higher versus adaptation to lower speed.}
     \label{fig:RCSP_filter}
\end{figure*}

Figure \ref{fig:LossVScomponents} and Figure \ref{fig:RCSP_filter} demonstrate the analysis of the most significant RCSP components and their spatial distribution, respectively. Figure \ref{fig:LossVScomponents} shows the average across subjects, 10-fold cross-validation, generalisation loss as we pick more RCSP components from the most significant to less significant eigen values. We note that for both type of two-class classification the error drops significantly for the first five components but it remains the same or even increases when more components are incorporated. 

Figure \ref{fig:RCSP_filter} shows the spatial distribution of the five more significant RCSP components for one of the subjects. The top row shows the RCSP filters associated with adaptation/non-adaptation classification, whereas the middle row shows the RCSP filters associated with right/left classification. The bottom row shows the spatial distribution of the filters associated with adaptation towards higher pace versus adaptation towards lower pace.

\section{Discussion}
Gait adaptation plays a significant role in the ability of humans to walk and maintain their balance. In elderly and people with neurological problems, it is an index of their health progression. Therefore, exoskeletons and assistive robotic devices should be able to sense and quickly adjust to gait changes \cite{Varghese2018}. This requires decoding neural signals accurately while people walk in their natural environments. Most of the adaptation studies today are based on specialized equipment such as split-zone treadmill, whereas they monitor gait with pressure insoles or reflective markers/multi-camera systems.
We have developed a framework to study gait adaptation in natural settings. The subjects walk in a room following the pace of a tone that changes between three modes of slow, normal and fast pace, randomly. We record EEG signal wirelessly whereas gait characteristics are extracted based on a single RGB camera. 

The EEG signal is preprocessed based on a bandpass filter of 3-45 Hz, followed by ICA to identify and remove motion-related artefacts. We use the extracted gait characteristics to epoch the EEG signal and to formulate three classification problems of intention detection in gait adaptation: i) right versus left step, ii) adaptation steps versus non-adaptation steps and iii) adaptation to higher pace versus adaptation to lower pace versus non-adaptation. Subsequently, we use RCSP to extract EEG features that maximize the discriminability between two classes.

RCSP is a regularized version of CSP algorithm that allows for automatic regularization of the covariance matrices for each class. The extent of regularization depends on the intrinsic properties of the data. Here we note a significant improvement of our results based on the RCSP filters for all types of classification. Furthermore, we show that incorporating both amplitude and frequency based characteristics can further improve the outcome of the classifier. 

CSP/RCSP is intrinsically a two-class feature extraction method. To extend it to three classes we extract features from three pairs of classes: i) adaptation to higher pace vs non-adaptation, ii) adaptation to lower pace vs non-adaptation and iii) adaptation to lower pace versus adaptation to higher pace.  

Finally, we investigate the influence of the number of components to the classification accuracy and their spatial distribution. We note that for the two-class experiments five out of 32 components are enough to achieve high accuracy. Furthermore, the spatial distribution of the RCSP components seems to be of physiological origin. 

\balance

\bibliography{IEEEtrans}
\end{document}